\documentclass[%
 reprint,
nofootinbib,
amsmath,
amssymb,
aps,
floatfix,
twocolumn
]{revtex4-2}

\usepackage{graphicx}
\usepackage{dcolumn}
\usepackage{bm}
\usepackage{booktabs}
\usepackage{braket} 
\usepackage{amsmath}
\usepackage{float}
\usepackage{multirow}
\usepackage{hyperref}
\usepackage[dvipsnames]{xcolor}
\usepackage[compat=1.1.0]{tikz-feynman}
\usepackage[compat=1.1.0]{tikz-feynhand}
\hypersetup{
    colorlinks,
    linkcolor={Emerald},
    citecolor={Emerald},
    urlcolor={Emerald}
}
\usepackage{tikz}
\usepackage[compat=1.1.0]{tikz-feynman}
\usepackage[compat=1.1.0]{tikz-feynhand}
\usepackage[export]{adjustbox}
\usepackage{color, colortbl}

\newcommand{\sig}{\sigma}
\newcommand{\lzm}{\left(}
\newcommand{\dzm}{\right)}
\newcommand{\lzs}{\left[}
\newcommand{\dzs}{\right]}

\newcommand{\lzu}{\left|}
\newcommand{\dzu}{\right|}
\newcommand{\cL}{\mathcal{L}}
\newcommand{\cO}{\mathcal{O}}

\newcommand{\cW}{{\mathcal W}}
\newcommand{\cB}{{\mathcal B}}

\newcommand{\cQ}{{\mathcal Q}}

\newcommand{\cG}{{\mathcal G}}

\newcommand{\cY}{{\mathcal Y}}
\newcommand{\cH}{{\mathcal H}}

\newcommand{\im}{{\mathrm{Im}} \,}
\newcommand{\hermc}{\text{h.c.}}

\newcommand{\Tr}{\mathop{\mathrm{Tr}}}

\newcommand{\sscript}[1]{{\scriptscriptstyle \mathrm{#1}}}

\begin{document}
\title{Leading directions in the SMEFT: Renormalization Effects}

\author{Admir Greljo}
 \email{admir.greljo@unibas.ch}
\author{Ajdin Palavri\'c}
 \email{ajdin.palavric@unibas.ch}
\author{Aleks Smolkovi\v c}%
 \email{aleks.smolkovic@unibas.ch}

\affiliation{%
 Department of Physics, University of Basel, Klingelbergstrasse 82,  CH-4056 Basel, Switzerland
}%

\begin{abstract}

The stability of the electroweak scale, challenged by the absence of deviations in flavor physics, prompts the consideration of SMEFT scenarios governed by approximate SM flavor symmetries. This study examines microscopic theories that match onto a set of $U(3)^5$-symmetric dimension-6 operators. Renormalization group mixing from the ultraviolet to the electroweak scale yields significant phenomenological constraints, particularly pronounced for UV-motivated directions. To demonstrate this, we explore a complete suite of tree-level models featuring new spin-0, 1/2, and 1 fields, categorized by their irreducible representations under the flavor group. We find that for the \textit{leading directions}, corresponding to a single-mediator dominance, RG mixing effects occasionally serve as the primary indirect probe.

\end{abstract}

\maketitle

\section{\label{sec:setup}Introduction}

Given a perceptible gap between the new physics (NP) scale and the electroweak (EW) scale, the Standard Model Effective Field Theory (SMEFT)~\cite{Buchmuller:1985jz, Grzadkowski:2010es, Brivio:2017vri, Isidori:2023pyp, Giudice:2007fh, Henning:2014wua} emerges as a robust theoretical framework for describing deviations from the Standard Model (SM). The SMEFT Lagrangian is an infinite series of higher-dimensional local operators built from the SM fields obeying gauge and Poincaré symmetries. The respective Wilson coefficients (WCs) encapsulate the short-distance effects of a broad spectrum of models beyond the SM (BSM). In the absence of a clear direction toward a specific BSM scenario, such a framework provides a convincing path forward, informing phenomenological studies and data interpretation.

The minimal number of independent WCs (an operator basis) is rendered finite at each order in the inverse powers of the cutoff scale controlled by the canonical dimensions. Yet, the size of this space rapidly increases with the growing canonical dimension, but also with the number of families~\cite{Henning:2015alf}. Specifically, for leading-order baryon-number conserving operators at dimension six, the parameter count rises from 59 for a single active generation to a striking 2499 for three generations~\cite{Alonso:2013hga}. This escalation underscores the complexity introduced by the flavor degrees of freedom.

On the other hand, the fermion kinetic terms enjoy a large $U(3)^5$ global symmetry owning to the three copies of five different gauge representations. The Yukawa interactions induce a rather peculiar explicit breaking, giving rise to exact and approximate flavor symmetries in the SM. The absence of violation of the implied selection rules in precision flavor experiments, such as $\Delta F=2$ transitions, charged lepton flavor violation, and electric dipole moments, already imposes stringent constraints on NP which does not lie far above the EW scale~\cite{EuropeanStrategyforParticlePhysicsPreparatoryGroup:2019qin}. Indeed, a viable TeV-scale physics, anticipated by the Higgs hierarchy problem and driving direct searches at the energy frontier, should not excessively violate the approximate flavor symmetries. This reasoning motivates the introduction of flavor power counting in the SMEFT, allowing for more focused analyses. Indeed, flavor symmetries prove to be very beneficial in charting the space of the SMEFT~\cite{Greljo:2022cah, Faroughy:2020ina}.

In this work, we consider a class of microscopic theories that integrate out to $U(3)^5$-symmetric dimension-6 basis made up of only 47 operators. These are the leading operators in the minimal flavor violation (MFV)~\cite{DAmbrosio:2002vsn} power counting and represent the most minimal complete operator basis of interest for global fits of top, Higgs, and electroweak data~\cite{Aoude:2020dwv, Bruggisser:2022rhb, Bartocci:2023nvp, Grunwald:2023nli}. As such, it constitutes an important initial playground towards more complicated global analyses, such as those based on the $U(2)^5$ flavor symmetries~\cite{Barbieri:2011ci, Kagan:2009bn, Fuentes-Martin:2019mun, Allwicher:2023shc}.\footnote{Smaller groups like $U(2)_{q+e}$ account for the peculiar fermion masses and mixings but offer limited protection against flavor constraints (see Fig.1 in \cite{Antusch:2023shi}). }

Restricting ourselves to $U(3)^5$-symmetric operators at the ultraviolet (UV) matching scale, the main focus of this investigation is on the renormalization group (RG) effects between the UV and the EW scales. These effects are governed by the SMEFT anomalous dimension matrix computed in~\cite{Alonso:2013hga, Jenkins:2013wua, Jenkins:2013zja} (see also~\cite{Machado:2022ozb}) and implemented in numerical tools~\cite{Dawson:2022ewj} such as \texttt{wilson}~\cite{Aebischer:2018bkb}, \texttt{DsixTools}~\cite{Celis:2017hod, Fuentes-Martin:2020zaz} and \texttt{RGESolver}~\cite{DiNoi:2022ejg}. While it has become a common practice to automatically include these effects, for example, in \texttt{smelli}~\cite{Aebischer:2018iyb}, our work aims to pinpoint the most constrained linear combinations of operators at the UV matching scale resulting solely from RG mixing. To deepen the understanding of these effects, we provide simplified analytical expressions supported by the full numerical results.
Our key interest lies in identifying RG-induced contributions to precision observables at low energies, which offer stronger or comparable bounds to those from tree-level processes. Upon examination, noteworthy cases involve four- and two-quark operators, where the RG bounds rival those from top quark production~\cite{Ethier:2021bye}, echoing recent findings for top-specific operators~\cite{Garosi:2023yxg}.

To demonstrate the significance of RG effects, we examine a full set of tree-level mediator models matching onto the $U(3)^5$-symmetric operator basis at the UV scale. A complete spectrum of mediator fields with spin 0, 1/2, and 1, along with their SM and $U(3)^5$ flavor representations, has been comprehensively identified and matched to the universal basis in Ref.~\cite{Greljo:2023adz}, building upon~\cite{deBlas:2017xtg}. This matching process has defined a finite set of \textit{leading directions} -- UV-motivated linear combinations of the WCs, warranting thorough examination. This paper performs a complete RG analysis of all these leading directions, going beyond the tree-level phenomenology presented in~\cite{Greljo:2023adz}. Our central findings are showcased in Tables~\ref{table:EFTresults_scalars}, \ref{table:EFTresults_vectors} and \ref{table:EFTresults_fermions}, comparing the RG bounds on a comprehensive set of four- and two-quark leading directions with the tree-level bounds.

This paper is structured as follows: in Section~\ref{sec:RGE_universal_SMEFT} we identify crucial RG equations, Section~\ref{sec:low_energy_constraints} gives an overview of the most sensitive low-energy observables, expressing them in terms of the WCs at the UV matching scale in the leading-log approximation. In Section~\ref{sec:applications}, we derive a comprehensive set of bounds on the leading directions, which are then compared against exclusions from direct searches for selected benchmark models. The summary and the future outlook are presented in Section~\ref{sec:conclusions}.

\section{RG effects from flavor-blind UV}
\label{sec:RGE_universal_SMEFT}

The full $U(3)^5$-symmetric operator basis is defined in Appendix~B of Ref.~\cite{Greljo:2023adz}. Focusing on the phenomenologically important RG effects, we consider a subset of operators at the UV matching scale involving quarks.\footnote{For all leading directions involving leptons, we have verified that tree-level effects consistently dominate (see Section~\ref{sec:applications}). The same is true for purely bosonic operators with the exception of $\cO_\phi$ already discussed in~\cite{Greljo:2023adz}.} Our starting point is the Lagrangian
\begin{equation}\label{eq:SMEFT_general}
    \cL_{\sscript{SMEFT}}\supset \sum_i C_i\cO_i\,,
\end{equation}
where the sum goes over all four-quark and two-quark-two-$\phi$ operators defined in Table~\ref{tab:MFV_basis} above the double line. Here, $q$ and $\ell$ denote the left-handed quark and lepton doublets, while $u$ and $d$ denote the right-handed up and down quark fields. $\phi$ denotes the Higgs doublet. Flavor indices are $i,j=1,2,3$, and the summation over repeated indices is assumed. In the rest of the paper, the labels assigned to the WCs, which we treat as dimensionful parameters, correspond directly to the labels of the operators from Table~\ref{tab:MFV_basis}.

Starting from Eq.~\eqref{eq:SMEFT_general} at the UV matching scale, the RG equations~\cite{Jenkins:2013zja, Jenkins:2013wua, Alonso:2013hga} determine the non-zero dimension-6 WCs at the EW scale, where the subsequent matching to the low-energy effective field theory (LEFT)~\cite{Jenkins:2017jig} is performed at the tree level.\footnote{See Ref.~\cite{Hurth:2019ula} for one-loop matching effects in the flavor-symmetric SMEFT.} There are three categories of important RG effects depicted in Figure~\ref{fig:diag_qq_phiq}:
\begin{itemize}
    \item \textbf{A}. Four-quark operator mixing into an EW boson vertex.
    \item \textbf{B}. Four-quark operator mixing with two insertions of Yukawa interactions.
    \item \textbf{C}. Four-quark operator mixing into a semileptonic operator. 
\end{itemize}

While gauge interactions are flavor-diagonal (FD), the Yukawa-dependent part of the anomalous dimension matrix leads to FD and flavor-violating (FV) effects. For example, when matched to the LEFT, the resulting operators from category \textbf{A} produce both FD and FV $Z$ couplings. The latter ones play a role in $\Delta F =1$ processes such as $b\to s \ell \ell$ decays.\footnote{We have verified that two insertions of FV $Z$ couplings, effectively suppressed by $1/\Lambda^4$, as well as beyond the leading-log mixing of four-quark operators, yield subleading constraints from $\Delta F=2$ processes compared to $\Delta F=1$.} 

\setlength{\tabcolsep}{2.2em} 
{\renewcommand{\arraystretch}{1.2}
\begin{table}[t]
    \centering
\scalebox{1}
{
	\begin{tabular}{cc}
		\toprule
		\textbf{Label} & \textbf{Operator}\\
		\midrule
		$\cO_{qq}^{(1)D}$&$(\bar q_{i}\gamma^\mu q^{i})(\bar q_{j}\gamma_\mu q^{j})$\\[2pt]
		$\cO_{qq}^{(3)D}$&$(\bar q_{i}\gamma^\mu \sig^a q^{i})(\bar q_{j}\gamma_\mu \sig^a q^{j})$\\[2pt]
		$\cO_{qq}^{(1)E}$&$(\bar q_{i}\gamma^\mu q^{j})(\bar q_{j}\gamma_\mu q^{i})$\\[2pt]
		$\cO_{qq}^{(3)E}$&$(\bar q_{i}\gamma^\mu \sig^a q^{j})(\bar q_{j}\gamma_\mu \sig^a q^{i})$\\[2pt]
		\midrule
		$\cO_{dd}^D$& $(\bar d_{i}\gamma^\mu d^{i})(\bar d_{j}\gamma_\mu d^{j})$\\[2pt]
		$\cO_{dd}^E$& $(\bar d_{i}\gamma^\mu d^{j})(\bar d_{j}\gamma_\mu d^{i})$\\[2pt]
		\midrule
		$\cO_{uu}^D$& $(\bar u_{i}\gamma^\mu u^{i})(\bar u_{j}\gamma_\mu u^{j})$\\[2pt]
		$\cO_{uu}^E$& $(\bar u_{i}\gamma^\mu u^{j})(\bar u_{j}\gamma_\mu u^{i})$\\[2pt]
		\midrule
		$\cO_{ud}^{(1)}$& $(\bar u_{i}\gamma^{\mu} u^{i})(\bar d_{j}\gamma_{\mu}  d^{j})$\\[2pt]
		$\cO_{ud}^{(8)}$& $(\bar u_{i}\gamma^{\mu}T^A u^{i})(\bar d_{j}\gamma_{\mu} T^A d^{j})$\\[2pt]
		\midrule
		$\cO_{qu}^{(1)}$& $(\bar q_{i}\gamma^\mu q^{i})(\bar u_{j}\gamma_\mu u^{j})$\\[2pt]
		$\cO_{qu}^{(8)}$& $(\bar q_{i}\gamma^\mu T^A q^{i})(\bar u_{j}\gamma_\mu T^A u^{j})$\\[2pt]
		$\cO_{qd}^{(1)}$& $(\bar q_{i}\gamma^\mu q^{i})(\bar d_{j}\gamma_\mu d^{j})$\\[2pt]
		$\cO_{qd}^{(8)}$& $(\bar q_{i}\gamma^\mu T^A q^{i})(\bar d_{j}\gamma_\mu T^A d^{j})$\\[2pt]
		\midrule
		$\cO_{\phi q}^{(1)}$& $(\phi^\dag i\overset{\text{\footnotesize$\leftrightarrow$}}{D}_\mu \phi)(\bar q_i\gamma^\mu q^i)$\\[2pt]
		$\cO_{\phi q}^{(3)}$& $(\phi^\dag i\overset{\text{\footnotesize$\leftrightarrow$}}{D^a_\mu} \phi)(\bar q_i\gamma^\mu\sig^a q^i)$\\[2pt]
		$\cO_{\phi u}$& $(\phi^\dag i \overset{\text{\footnotesize$\leftrightarrow$}}{D}_\mu \phi)(\bar u_i\gamma^\mu u^i)$\\[2pt]
		$\cO_{\phi d}$& $(\phi^\dag i\overset{\text{\footnotesize$\leftrightarrow$}}{D}_\mu \phi)(\bar d_i\gamma^\mu d^i)$\\[2pt]
		\midrule\midrule
		$\cO_{\ell q}^{(3)}$&$(\bar\ell_{i}\gamma^\mu\sig^a \ell^{i})(\bar q_{j}\gamma_\mu\sig^a q^{j})$\\[2pt]
        $\cO_{\ell\ell}^E$&$(\bar\ell_{i}\gamma^\mu\ell^{j})(\bar\ell_{j}\gamma_\mu\ell^{i})$\\[2pt]
		$\cO_{\phi\ell}^{(3)}$ & $(\phi^\dag i\overset{\text{\footnotesize$\leftrightarrow$}}{D^a_\mu} \phi)(\bar\ell_i\gamma^\mu\sig^a\ell^i)$\\[2pt]
		$\cO_{\phi D}$& $(\phi^\dag D_\mu \phi)[(D^\mu\phi)^\dag\phi]$\\[2pt]
		\bottomrule
	\end{tabular}
}
	\caption{$U(3)^5$-symmetric dimension-6 SMEFT operators appearing at the UV matching scale in this work are listed above the double line. Other operators enter through the RG mixing. Our notation closely follows Ref.~\cite{Greljo:2023adz}. All operators considered are Hermitian, ensuring that their corresponding WCs are real.} \label{tab:MFV_basis}
\end{table}

\subsection{Vertex corrections}
\label{sec:vertex-cor}

\begin{figure}
\begin{align}
&\textbf{A}. &\quad\quad &\begin{tikzpicture}[baseline=(a1.base), scale=0.65, transform shape]
\begin{feynman}
\vertex (a1) {\(\)};
\vertex [square dot][left = 1.1cm of a1](a2) {\(\)};
\vertex [left = 1.1cm of a2](a11) {\LARGE\(C_{qq}^{(1)}\)};
\vertex [above =1.3cm of a2] (a3) {\LARGE\(q\)};
\vertex [below =1.3cm of a2] (a4) {\LARGE\(q\)};
\vertex [left =1.7cm of a3] (a5) {\LARGE\(q\)};
\vertex [left =1.7cm of a4] (a6) {\LARGE\(q\)};
\vertex [above =1.25cm of a1] (a7);
\vertex [above =0.5cm of a7] (a7771);
\vertex [right =3cm of a7771] (a77771);
\vertex [above =0.2cm of a7] (a77) {\LARGE\(Y_{u}\)};
\vertex [below =1.25cm of a1] (a8);
\vertex [below =0.5cm of a8] (a888);
\vertex [right =3cm of a888] (a8881);
\vertex [below =0.2cm of a8] (a88) {\LARGE\(Y_{u}^\dag\)};
\vertex [right =1.5cm of a7] (a9);
\vertex [right =1.5cm of a9] (a11);
\vertex [below =1.5cm of a9] (a10);
\vertex [left =-0.5cm of a10] (a100) {\LARGE\(u\)};
\vertex [right =1.5cm of a100] (a101);
\vertex [above =1.9cm of a101] (a102) {\LARGE\(\phi\)};
\vertex [below =1.5cm of a101] (a103) {\LARGE\(\phi\)};
\diagram*{
(a5)--[plain](a2)-- [plain](a6)--[plain] (a2),
(a7)--[plain, half right](a8),
(a7)--[plain, half left](a8),
(a7)--[scalar](a102),
(a8)--[scalar](a103)
};
\draw[fill=black] (a7) circle(1.0mm);
\draw[fill=black] (a8) circle(1.0mm);
\end{feynman}
\end{tikzpicture} \nonumber\\
&\textbf{B}. &\quad\quad &\begin{tikzpicture}[baseline=(a1.base), scale=0.65, transform shape]
\begin{feynman}
\vertex (a1) {\(\)};
\vertex [square dot][left = 1.1cm of a1](a2) {\(\)};
\vertex [left = 1.1cm of a2](a11) {\LARGE\(C_{uu}\)};
\vertex [above =1.3cm of a2] (a3) {\LARGE\(u\)};
\vertex [below =1.3cm of a2] (a4) {\LARGE\(u\)};
\vertex [left =1.7cm of a3] (a5) {\LARGE\(u\)};
\vertex [left =1.7cm of a4] (a6) {\LARGE\(u\)};
\vertex [above =1.25cm of a1] (a7);
\vertex [above =0.5cm of a7] (a7771);
\vertex [right =3cm of a7771] (a77771);
\vertex [above =0.2cm of a7] (a77) {\LARGE\(Y_{u}\)};
\vertex [below =1.25cm of a1] (a8);
\vertex [below =0.5cm of a8] (a888);
\vertex [right =3cm of a888] (a8881);
\vertex [below =0.2cm of a8] (a88) {\LARGE\(Y_{u}^\dag\)};
\vertex [right =1.5cm of a7] (a9);
\vertex [right =1.5cm of a9] (a11);
\vertex [below =1.5cm of a9] (a10);
\vertex [left =-0.3cm of a10] (a100) {\LARGE\(\phi\)};
\vertex [right =1.5cm of a100] (a101);
\vertex [above =1.9cm of a101] (a102) {\LARGE\(q\)};
\vertex [below =1.5cm of a101] (a103) {\LARGE\(q\)};
\diagram*{
(a5)--[plain](a2)-- [plain](a6)--[plain] (a2),
(a7)--[plain, half right](a8),
(a7)--[scalar, half left](a8),
(a7)--[plain](a102),
(a8)--[plain](a103)
};
\draw[fill=black] (a7) circle(1.0mm);
\draw[fill=black] (a8) circle(1.0mm);
\end{feynman}
\end{tikzpicture}\nonumber\\
&\textbf{C}. &\quad &\begin{tikzpicture}[baseline=(a1.base), scale=0.65, transform shape]
\begin{feynman}
\vertex (a1);
\vertex [square dot][left = 1.1cm of a1](a2) {\(\)};
\vertex [left = 0.9cm of a2](a11) {\LARGE\(C_{qq}^{(3)}\)};
\vertex [above = 1.1cm of a1](a15) {\LARGE\(q\)};
\vertex [below = 1.1cm of a1](a16) {\LARGE\(q\)};
\vertex [right = 1.1cm of a1](a3);
\vertex [left = 1.5cm of a2](a4);
\vertex [above = 1.5cm of a4](a5) {\LARGE\(q\)};
\vertex [below = 1.5cm of a4](a6) {\LARGE\(q\)};
\vertex [right = 1.8cm of a3](a7);
\vertex [right = 1.8cm of a7](a8);
\vertex [above = 1.5cm of a8](a9) {\LARGE\(\ell\)};
\vertex [below = 1.5cm of a8](a10) {\LARGE\(\ell\)};
\diagram*{
(a2)--[plain, half left](a3)--[plain, half left](a2),
(a6)--[plain](a2)--[plain](a5),
(a3)--[photon, edge label=\LARGE\({W}\)](a7),
(a9)--[plain](a7)--[plain](a10)};
\draw[fill=black] (a7) circle(1.0mm);
\draw[fill=black] (a3) circle(1.0mm);
\end{feynman}
\end{tikzpicture}\nonumber
\end{align}
\caption{Representative RG diagrams. See Section~\ref{sec:RGE_universal_SMEFT} for details. }
\label{fig:diag_qq_phiq}
\end{figure}
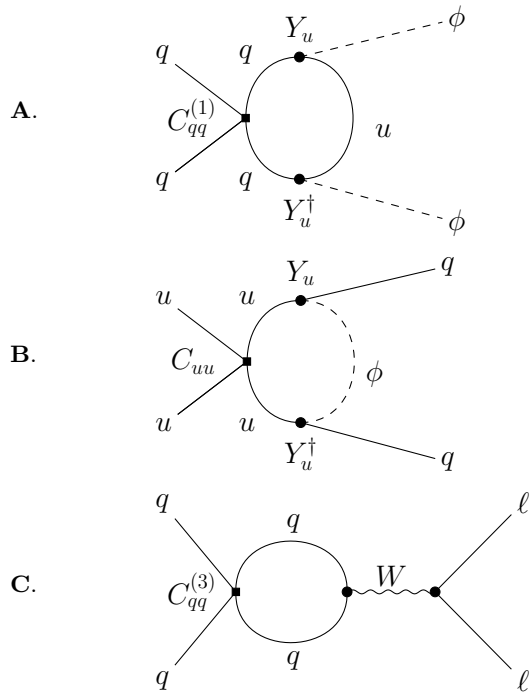

As depicted by the top diagram in Figure~\ref{fig:diag_qq_phiq}, operators $\mathcal{O}_{qq}^{(1,3)}$ mix into $\mathcal{O}_{\phi q}^{(1,3)}$ by closing the loop with an up-type quark and emitting two Higgs fields. See Ref.~\cite{Jenkins:2013wua} for the flavor-generic RG expressions. These contributions are proportional to $Y_u Y_u^\dag$, leading to $y_t^2$-enhanced effects.\footnote{Throughout this work, we will not discuss small contributions $\propto Y_d Y_d^\dagger $, although they are included in our numerical studies.}

All such effects, including the running of the operators from the $\psi^2 H^2 D$ class, can be described by the following set of RG equations
\begin{equation}
\label{eq:rge:4q_into_VC}
	\begin{alignedat}{2}
		\dot C_{\underset{pr}{\phi q}}^{(1)} &=C_{\phi q, \sscript{FV}}^{(1)} [Y_uY_u^\dag]_{pr}+C_{\phi q, \sscript{FD}}^{(1)}\delta_{pr}\,,
		\\
		\dot C_{\underset{pr}{\phi q}}^{(3)} &=C_{\phi q, \sscript{FV}}^{(3)}[Y_uY_u^\dag]_{pr}+C_{\phi q, \sscript{FD}}^{(3)}\delta_{pr}\,,
		\\
		\dot C_{\underset{pr}{\phi u}} &=C_{\phi u, \sscript{FD}}^{1}[Y_u^\dag Y_u]_{pr}+C_{\phi u, \sscript{FD}}^{2}\delta_{pr}\,,
		\\
		\dot C_{\underset{pr}{\phi d}} &=C_{\phi d, \sscript{FD}}\delta_{pr}\,,
	\end{alignedat}
\end{equation}
where the abbreviation of the form $\dot{C}\equiv 16 \pi^2 \mu \frac{d}{d\mu} C$ is used, and we introduce the linear combinations of WCs in the flavor-symmetric basis\footnote{We omit contributions proportional to $g_1$ and smaller parameters, although they are fully included in the numerical analyses of Section~\ref{sec:applications}.}
	\begin{alignat}{2}
		C_{\phi q, \sscript{FV}}^{(1)}&\equiv2C_{qq}^{(1)D}+6C_{qq}^{(3)D}+12C_{qq}^{(1)E}
		-C_{\phi u}
		\nonumber
		\\&
		\phantom{....}+4C_{\phi q}^{(1)}-9C_{\phi q}^{(3)}
		\,,
		\nonumber\\
		C_{\phi q, \sscript{FD}}^{(1)}&\equiv \Big(2C_{qq}^{(1)E}+6C_{qq}^{(3)E}+12C_{qq}^{(1)D}
		\nonumber-6 C_{qu}^{(1)}\\&
		\phantom{....}+6C_{\phi q}^{(1)}\Big) y_t^2
		\nonumber
		\,,
		\nonumber\\
		C_{\phi q, \sscript{FV}}^{(3)}&\equiv -2C_{qq}^{(1)D}+2C_{qq}^{(3)D}-12C_{qq}^{(3)E}-3C_{\phi q}^{(1)}\nonumber\\&
		\phantom{....}+2C_{\phi q}^{(3)}
        \,,
		\nonumber\\
		C_{\phi q, \sscript{FD}}^{(3)}&\equiv-2 \lzm C_{qq}^{(1) E}-C_{qq}^{(3)E}+6C_{qq}^{(3)D}-3C_{\phi q}^{(3)} \dzm y_t^2\nonumber
        \\
        &\phantom{....}+\frac{g_2^2}{3}\Big( 2C_{qq}^{(1)D}+6C_{qq}^{(1)E}+34C_{qq}^{(3)D}
        \nonumber+6C_{qq}^{(3)E}\\&
        \phantom{....}+C_{\phi q}^{(3)} \Big)\,,\nonumber
        \\
		C_{\phi u, \sscript{FD}}^{1}&\equiv -12 C_{uu}^E-4C_{uu}^D-2C_{\phi q}^{(1)}+8C_{\phi u}\,,
		\nonumber\\
		C_{\phi u, \sscript{FD}}^{2}&\equiv 2\lzm 3 C_{qu}^{(1)}-6C_{uu}^D-2C_{uu}^E +3C_{\phi u} \dzm y_t^2\,,
		\nonumber\\
		C_{\phi d, \sscript{FD}}&\equiv6\lzm C_{qd}^{(1)}-C_{ud}^{(1)}+C_{\phi d} \dzm y_t^2\,.
	\end{alignat}
When the operators $\mathcal{O}_{\phi q}^{(1,3)}$, $\mathcal{O}_{\phi u}$ and $\mathcal{O}_{\phi d}$, generated via Eq.~\eqref{eq:rge:4q_into_VC}, are matched to the LEFT, they induce modified $Z$ and $W$ couplings to quarks~\cite{Jenkins:2017jig}, contributing to important FD, as well as FV observables, see Section~\ref{sec:low_energy_constraints}.

\subsection{Four-quark operators}
\label{sec:4qops}

The four-quark operators given in Table~\ref{tab:MFV_basis} mix among themselves under RG equations, leading to interesting $\Delta F=1$ FV effects as illustrated by the middle diagram in Figure~\ref{fig:diag_qq_phiq}. The system of equations, simplified to include only $Y_u$-dependent terms, is as follows:
\begin{equation}
\label{eq:rge:4qmix}
	\begin{alignedat}{4}
		\dot C^{(1)}_{\underset{prst}{qu}} &= C_{qu,\sscript{FV}}^{(1)} [Y_uY_u^\dag]_{pr}\delta_{st}\,,
		&\quad 
		\dot C^{(1)}_{\underset{prst}{qd}} &=C_{qd,\sscript{FV}}^{(1)}[Y_uY_u^\dag]_{pr}\delta_{st}\,,
		\\ 
		\dot C^{(8)}_{\underset{prst}{qu}} &=C_{qu,\sscript{FV}}^{(8)} [Y_uY_u^\dag]_{pr}\delta_{st}\,,
		&\quad 
        \dot C^{(8)}_{\underset{prst}{qd}} &=C_{qd,\sscript{FV}}^{(8)} [Y_uY_u^\dag]_{pr}\delta_{st}\,,
	\end{alignedat}
\end{equation}
where we introduce the linear combinations
\begin{equation}
	\begin{alignedat}{4}
		C_{qu,\sscript{FV}}^{(1)}&\equiv C_{qu}^{(1)}-\frac{2}{3}C_{uu}^E-2C_{uu}^D\,,
		&\quad 
		C_{qd,\sscript{FV}}^{(1)}&\equiv C_{qd}^{(1)}-C_{ud}^{(1)}\,,
		\\
		C_{qu,\sscript{FV}}^{(8)}&\equiv C_{qu}^{(8)}-4C_{uu}^E\,,
		&\quad 
		C_{qd,\sscript{FV}}^{(8)}&\equiv C_{qd}^{(8)}-C_{ud}^{(8)}\,.
	\end{alignedat}
\end{equation}
As in previous cases, these are formulated in terms of the WCs of operators from the flavor-symmetric basis. The RG equations presented here are not exhaustive but capture only the phenomenologically relevant terms.

\subsection{Semileptonic operators}

For the upcoming analysis of the low-energy observables, as it unfolds, keeping only the RG terms proportional to $g_2^2$, while neglecting the ones proportional to $g_1^2$, there is only one relevant RG-generated semileptonic operator, which appears as a result of the $\mathcal{O}_{qq}^{(3)}$ mixing into $\mathcal{O}_{\ell q}^{(3)}$~\cite{Alonso:2013hga}. The RG equation for $\mathcal{O}_{\ell q}^{(3)}$ operator takes the form
\begin{equation}
\label{eq:rge:4q_into_lq}
		\dot C_{\underset{prst}{\ell q}}^{(3)}=g_2^2\,C_{\ell q,\sscript{FD}}^{(3)}\delta_{pr}\delta_{st}\,,
\end{equation}
where we introduce
\begin{equation}
		C_{\ell q,\sscript{FD}}^{(3)}\equiv
		\frac{2}{3}\lzm C_{qq}^{(1)D}+3C_{qq}^{(1)E}+17C_{qq}^{(3)D}+3C_{qq}^{(3)E} \dzm\,.
\end{equation}
To reemphasize, our numerical analysis in Section~\ref{sec:applications} does not employ such approximations. Nonetheless, the above equations effectively approximate the most sensitive RG effects.

\section{RG-induced low-energy probes}\label{sec:low_energy_constraints}

Having detailed the relevant RG equations, this section focuses on their impact on key low-energy probes. We address this by solving these equations using a leading-log approximation, which allows us to express low-energy (pseudo)-observables in terms of the WCs at the UV matching scale.

\subsection{$b\to s \ell \ell$}

Flavor-violating $Z$ couplings to quarks, represented by the $\mathcal{O}_{\phi q}^{(1,3)}$ matching to LEFT, are effectively constrained from rare meson decays to charged leptons or neutrinos.\footnote{Down-quark FV is absent when $C_{\phi q}^{(1)}=-C_{\phi q}^{(3)}$ at the EW scale.} Presently, available data from charged leptons is more constraining, while final states with neutrinos can provide a complementary test in the future. Furthermore, correlated effects are predicted in all down-quark FV neutral currents, including $b\to s$, $b\to d$, and $s \to d$ transitions. In fact, the measurements of $b\to s \ell \ell$ decays provide the most sensitive probe of semileptonic interactions with MFV structure, see Ref.~\cite{Greljo:2022jac}.

Rare $b$ decays are described by the weak Hamiltonian
\begin{equation}
\cH_{\sscript{eff.}} \supset -\frac{4G_F}{\sqrt2}\frac{\alpha}{4\pi} V_{ts}^*V_{tb}
	\lzm C_{9}\,\cO_{9}+C_{10}\,\cO_{10}\dzm +\hermc\,,
\end{equation}
where $G_F$ is the Fermi constant and $\alpha$ is the fine-structure constant, while the local operators are defined as
\begin{equation}
    \cO_9=(\bar\ell\gamma_\mu \ell)(\bar s_L\gamma^\mu b_L)\,,
    \quad 
    \cO_{10}=(\bar\ell\gamma_\mu\gamma_5 \ell)(\bar s_L\gamma^\mu b_L)\,,
\end{equation}
and $C_{9,10}$ denote the short-distance contributions. NP contributions through modified $Z$ couplings to quarks predict lepton flavor universality. In addition, $ C^{{\sscript{NP}}}_9 $ is suppressed due to the small $Z$ couplings to the leptonic vector current. Conversely, $C^{{\sscript{NP}}}_{10}$ receives significant NP contributions. 

After solving the RG equations~\eqref{eq:rge:4q_into_VC} in the leading-log approximation and matching SMEFT onto LEFT at the tree level, we find
\begin{equation}
\label{eq:C10_general_formula}
	C^{{\sscript{NP}}}_{10}=-\frac{v^2}{4e^2}y_t^2\lzm C^{(1)}_{\phi q,\sscript{FV}}+C^{(3)}_{\phi q,\sscript{FV}} \dzm\ln\lzm\frac{\mu_i}{\mu_f}\dzm\,,
\end{equation}
where $v\approx 246$\,GeV is the Higgs vacuum expectation value, $e$ is the elementary charge, $\mu_f=m_Z$ is the EW matching scale, and $\mu_i=\mathcal{O}(\mathrm{TeV})$ is the UV matching scale.\footnote{Note, that finite one-loop matching contributions are subleading when compared to large log-enhanced RG effects~\cite{Hurth:2019ula, Aoude:2020dwv}.} In the following, we will use $C^{{\sscript{NP}}}_{10}$ as a pseudo-observable, contrasting it with the best-fit interval from global fits to $b \to s \ell \ell$ data, $C^{{\sscript{NP}}}_{10} = 0.23\pm 0.15$~\cite{Greljo:2022jac}, see also~\cite{Alguero:2023jeh, Ciuchini:2022wbq, Hurth:2023jwr}.

\subsection{$\varepsilon^\prime/\varepsilon$}
The $\varepsilon^\prime/\varepsilon$ ratio measures the size of direct ($\Delta F=1$) CP violation in $K_L\to \pi \pi$ relative to indirect ($\Delta F=2$) CPV. The current experimental world average is $(\epsilon^\prime/\epsilon)_\mathrm{exp} = (16.6\pm2.3)\times 10^{-4}$~\cite{ParticleDataGroup:2022pth}. As for the SM prediction, we take the current best estimate as $(\epsilon^\prime/\epsilon)_\mathrm{SM} = (13.9\pm5.2)\times 10^{-4}$~\cite{Buras:2020pjp, Aebischer:2020jto, Aebischer:2021hws}.

In our framework, contributions to this observable are generated by the RG mixing of the four-quark operators (Section~\ref{sec:4qops}), along with the four-quark operators mixing into the gauge boson vertex corrections (Section~\ref{sec:vertex-cor}). Solving the RG equations~\eqref{eq:rge:4q_into_VC} and \eqref{eq:rge:4qmix} in the leading-log approximation, then matching the SMEFT onto the JMS LEFT basis \cite{Jenkins:2017jig}, and using the master formula provided in~\cite{Aebischer:2021hws,Aebischer:2018quc}, we obtain
\begin{alignat}{2}
\label{eq:eps'/eps}
		\lzm\varepsilon'/\varepsilon\dzm_{\sscript{BSM}}
		&=-\mathcal N_{\Delta S=1}\frac{y_t^2}{16\pi^2} \ln
        \lzm\frac{\mu_i}{\mu_f}\dzm\im[V_{ts}^*V_{td}]
        \nonumber\\
        &\times
        \Bigg\{
		\lzs C_{qu,\sscript{FV}}^{(1)}-\frac{4}{3}s_\theta^2\lzm C_{\phi q,\sscript{FV}}^{(1)}+C_{\phi q,\sscript{FV}}^{(3)} \dzm \dzs  P_{du}^{(1)}
		\nonumber\\
        &\phantom{-.}
		+\lzs C_{qd,\sscript{FV}}^{(1)}+\frac{2}{3}s_\theta^2\lzm C_{\phi q,\sscript{FV}}^{(1)}+C_{\phi q,\sscript{FV}}^{(3)} \dzm \dzs P_{dd}^{(1)}
		\nonumber\\
        &\phantom{.-}   +C_{qu,\sscript{FV}}^{(8)}P_{du}^{(8)}+C_{qd,\sscript{FV}}^{(8)}P_{dd}^{(8)}\Bigg\}\,,
\end{alignat}
where $\mathcal N_{\Delta S=1}=(1\,\mathrm{TeV})^{2}$. We only keep the contributions from the largest hadronic matrix elements, $P_{dd}^{(1,8)} \equiv P[C^{V(1,8),LR}_{\underset{2111}{dd}}]$, $P_{du}^{(1,8)} \equiv P[C^{V(1,8),LR}_{\underset{2111}{du}}]$~\cite{Aebischer:2021hws}, which capture all of the relevant effects in this study. Note that, in accordance with the aforementioned references, $\mu_f = 160~\mathrm{GeV}$ is used. As a final comment, although we start with strictly real WCs at the UV scale (see e.g.~Ref.~\cite{Fajfer:2023gie} for a discussion regarding imaginary WCs), through the process of RG mixing and matching, we end up with $\im[V_{ts}^*V_{td}]$, which then enters the CP-violating observable $\varepsilon^\prime/\varepsilon$.\footnote{This effect is independent of the flavor basis used. In the up basis, two CKM matrix element insertions arise when rotating $d$ quarks to mass eigenstates, while in the down basis, they emerge from rotating the $Y_u$ matrix in the RG equations.}

\subsection{$W$ mass}
\label{sec:mW}

We choose $(G_F, m_Z, \alpha)$ as an input parameter set and can therefore predict the value of $m_W$ both in the SM, including up to two-loop corrections~\cite{Awramik:2003rn}, and in $U(3)^5$-symmetric SMEFT~\cite{Berthier:2015oma, Bjorn:2016zlr, Bagnaschi:2022whn}. We express the combined prediction as $m_W^2 = m_{W,\sscript{SM}}^2  + \delta m_W^2$ with
\begin{equation}\label{eq:mW_correction}
    \frac{\delta m_W^2}{m_W^2} = - \frac{v^2 s_{2\theta}}{4c_{2\theta}} \lzs \frac{c_\theta}{s_\theta} C_{\phi D} + \frac{s_\theta}{c_{\theta}}(4 C^{(3)}_{\phi l} -2C^E_{ll})\dzs ,
\end{equation}
where $\theta$ is the Weinberg angle, $s_x \equiv \sin x$, and $c_x \equiv \cos x$.\footnote{It should be noted that Eq.~\eqref{eq:mW_correction} generally includes an additional term proportional to $C_{\phi WB}$~\cite{Berthier:2015oma, Bjorn:2016zlr, Bagnaschi:2022whn}.  However, operators from Table~\ref{tab:MFV_basis} do not mix into this operator at the one-loop level.}

The operator class $\psi^2H^2D^2$ from the $U(3)^5$-symmetric basis (where $\psi$ denotes quark fields) contributes to $m_W$ already at the leading log, by mixing into $\cO_{\phi D}$ and $\cO_{\phi\ell}^{(3)}$. The corresponding RG equations are given as
\begin{equation}
		\dot C_{\phi D}=24y_t^2\lzm C_{\phi q}^{(1)}-C_{\phi u} \dzm\,, 
        \quad
		\dot C_{\underset{rs}{\phi\ell}}^{(3)}=6g_2^2 C_{\phi q}^{(3)}\delta_{rs}\,.
\end{equation}
Solving these equations, we find
\begin{equation}
    \small
	\begin{alignedat}{2}
    \frac{\delta m_W^2}{m_W^2} &= \frac{3v^2}{8\pi^2}\frac{s_{2\theta}}{c_{2\theta}}\ln
    \lzm\frac{\mu_i}{\mu_f}\dzm
    \Bigg[y_t^2\frac{c_\theta}{s_\theta}\lzm C_{\phi q}^{(1)}-C_{\phi u} \dzm
    +g_2^2\frac{s_\theta}{c_\theta}C_{\phi q}^{(3)}
    \Bigg]\,.
	\end{alignedat}
\end{equation}
Interestingly, the four-quark operators considered in this work only contribute to $\delta m_W^2$ beyond the leading-log approximation. Specifically, all four-quark operators in Table~\ref{tab:MFV_basis} except color octets contribute at this order.\footnote{Color octets are absent because loops with color octet operator insertions in the relevant mixing cascades are always proportional to $\Tr(T^A)$, which is identically zero.} For example, $\cO(Y_u^4)$ contribution proceeds through $C_{uu}^{D,E}\to C_{\phi u}\to C_{\phi D}$ and similarly at $\cO(g_2^4)$, we have $C_{qq}^{(3)D,E}\to C_{\ell q}^{(3)}\to C_{\ell\ell}^{D,E}$. Here, we rely on a fully numerical solution of the RG equations using \texttt{wilson}~\cite{Aebischer:2018bkb}.

We use \texttt{flavio}~\cite{Straub:2018kue} to obtain the SM prediction, $m_{W}^{\sscript{SM}}= (80.355\pm 0.005)~\mathrm{GeV}$. As for the measured value, we consider the latest PDG combination of $m_{W}^{\sscript{exp.}} =~(80.377\pm0.012)~\mathrm{GeV}$~\cite{ParticleDataGroup:2022pth}, which does not include the anomalous CDF II result~\cite{CDF:2022hxs}.

\subsection{$Z$ pole observables}

As discussed in Section~\ref{sec:vertex-cor}, the operators considered in this work can lead to modifications of the FD $Z$-boson couplings with quarks. Such effects can be constrained from electroweak precision tests (EWPT) with on-shell $Z$ bosons~\cite{Efrati:2015eaa, deBlas:2015aea}. These constitute various partial decay width ratios, forward-backward asymmetries, and left-right asymmetries~\cite{ALEPH:2005ab, Efrati:2015eaa}. We utilize an extensive list of observables with correlated experimental errors, employing \texttt{smelli}~\cite{Aebischer:2018iyb} to build a custom EWPT likelihood focused solely on $Z$-pole observables. We exclude $m_W$ (analyzed separately in Section~\ref{sec:mW}) and $W$-pole observables sensitive to modified $W$ couplings to SM fermions, as they are not phenomenologically competitive. The constructed likelihood relies crucially on \texttt{flavio}~\cite{Straub:2018kue} due to its database of experimental measurements~\cite{SLD:2000jop, Janot:2019oyi, ALEPH:2005ab, ParticleDataGroup:2022pth} and the implemented theoretical predictions of the considered observables at the scale $\mu=m_Z$, including SM and BSM contributions~\cite{Freitas:2014hra, Brivio:2017vri}. In the subsequent numerical analysis, we use \texttt{wilson}~\cite{Aebischer:2018bkb} to run and match the WCs, capturing also beyond leading-log effects.

\subsection{$\beta$-decays}

Following the discussion in Section~\ref{sec:RGE_universal_SMEFT}, the only RG contributions to charged-current processes in our framework go through either modified $W$ couplings with left-handed quarks due to $\mathcal{O}_{\phi q}^{(3)}$, see Eq.~\eqref{eq:rge:4q_into_VC}, or through the $\mathcal{O}_{\ell q}^{(3)}$ contact interaction, see Eq.~\eqref{eq:rge:4q_into_lq}. Both of these match onto the same low-energy $V-A$ operator. The low-energy Hamiltonian is
\begin{equation}
\label{eq:Heff:cc}
    \mathcal{H}_{\sscript{eff}} \supset \frac{4 G_F}{\sqrt{2}} \sum_{x=d,s,b} \tilde{V}_{ux}\left(\bar{u}_L \gamma_\mu x_L\right) \left(\bar{e}_L \gamma_\mu \nu_{eL}\right)+\hermc\,.
\end{equation}
The NP contributions to the left-handed currents have been absorbed into $\tilde{V}_{ux}$ as $\tilde{V}_{ux} = V_{ux} (1+\epsilon_L^{x})$, 
where $V_{ux}$ are elements of the unitary rotation matrix~\cite{Falkowski:2017pss, Gonzalez-Alonso:2018omy}. The effects of nonzero $\epsilon_L^{x}$ can be probed through the violation of the (first row) CKM unitarity
\begin{equation}
\Delta_{\sscript{CKM}} \equiv |\tilde{V}_{ud}|^2 + |\tilde{V}_{us}|^2 + |\tilde{V}_{ub}|^2- 1 \,.
\end{equation}
Solving the respective RG equations in the leading-log approximation and matching onto Eq.~\eqref{eq:Heff:cc}, we obtain
\begin{equation}
    \epsilon_L^x=\frac{v^2}{16\pi^2}\left(g_2^2 C_{\ell q, \sscript{FD}}^{(3)}-C_{\phi q,\sscript{FD}}^{(3)}  \right)\ln\lzm\frac{\mu_i}{\mu_f}\dzm\,,
\end{equation}
for all $x=d,s,b$. The CKM unitarity test then reduces to $\Delta_\mathrm{CKM} \approx 2\epsilon_L^x$ at linear order in WCs, with no summation over $x$. 

The most accurate value of $|\tilde{V}_{ud}|$ is from superallowed $\beta$ decays, taken as the latest PDG average of $(0.97373\pm0.00031$~\cite{Hardy:2020qwl,ParticleDataGroup:2022pth}. For $|\tilde{V}_{us}|$, we use the conservative PDG average of semileptonic kaon decays and the kaon-to-pion decay ratio, at $0.2243\pm0.0008$~\cite{ParticleDataGroup:2022pth}. Including the minor $|\tilde{V}_{ub}|=(3.82\pm0.20)\times 10^{-3}$ contribution, the experimental constraint is $\Delta_{\sscript{CKM}}^{\sscript{exp.}}=(-1.52\pm0.70)\times 10^{-3}$, reflecting the known Cabibbo angle anomaly~\cite{Crivellin:2021bkd,Alok:2021ydy,Crivellin:2020ebi,Cirigliano:2023nol,Kirk:2020wdk}.

\subsection{Atomic parity violation}

Atomic parity violation (APV) is sensitive to parity-violating couplings of electrons to quarks. Experiments report the weak charge, defined as
\begin{equation}\label{eq:APV_general}
	Q_{\sscript{W}}(Z,N)=-2\lzs \lzm2Z+N\dzm g^{eu}_{\sscript{AV}}+\lzm Z+2N\dzm g^{ed}_{\sscript{AV}} \dzs\,,
\end{equation}
where $Z$ ($N$) is the number of protons (neutrons). The general expressions for $g^{eu}_{\sscript{AV}}$ and $g^{ed}_{\sscript{AV}}$ are given in~\cite{Falkowski:2017pss}. Upon solving the RG equations~\eqref{eq:rge:4q_into_VC} and \eqref{eq:rge:4q_into_lq} in the leading-log approximation, we get
\begin{equation}\label{eq:APV_g_quantities}
	\begin{alignedat}{2}
		g^{eu}_{\sscript{AV}}&\equiv-\frac{1}{2}+\frac{4}{3}s_\theta^2
		+\frac{v^2}{32\pi^2}\ln\lzm\frac{\mu_i}{\mu_f}\dzm
		\\&
		\phantom{.......}\times \bigg\{-g_2^2C_{lq,\sscript{FD}}^{(3)}-C_{\phi u,\sscript{FD}}^2
		+C_{\phi q,\sscript{FD}}^{(3)}-C_{\phi q,\sscript{FD}}^{(1)}
		\bigg\}\,,
		\\
		g^{ed}_{\sscript{AV}}&\equiv\frac{1}{2}-\frac{2}{3}s_\theta^2+\frac{v^2}{32\pi^2}\ln\lzm\frac{\mu_i}{\mu_f}\dzm\\&
		\phantom{.......}\times\bigg\{g_2^2C_{lq,\sscript{FD}}^{(3)}-C_{\phi d,\sscript{FD}}
		-C_{\phi q,\sscript{FD}}^{(3)}-C_{\phi q,\sscript{FD}}^{(1)}
		\bigg\}\,.
	\end{alignedat}
\end{equation}

The most precise measurements are done with the cesium ($^{133}\mathrm{Cs}$) atom, for which
\begin{equation}
Q_{\sscript{W}}^{\sscript{Cs}}\approx-376g^{eu}_{\sscript{AV}}-422g^{ed}_{\sscript{AV}}\,,
\end{equation}
up to small radiative corrections~\cite{ParticleDataGroup:2016lqr,ParticleDataGroup:2020ssz,Erler:2013xha}. The prediction for the cesium weak charge obtained in the SM is $Q_{\sscript{W}}^{\sscript{Cs},\sscript{SM}}=-73.23(1)$~\cite{Erler:2013xha,ParticleDataGroup:2020ssz,Cadeddu:2021dqx} 
which includes radiative corrections, while the current experimental value as reported by PDG is $Q_{\sscript{W}}^{\sscript{Cs},\sscript{exp.}}=-72.82(42)$~\cite{ParticleDataGroup:2020ssz,Wood:1997zq,Guena:2004sq}.

\section{Leading directions}
\label{sec:applications}

\setlength{\tabcolsep}{0.4em} 
{\renewcommand{\arraystretch}{1.4}
\begin{table*}
\centering
\scalebox{0.83}{
\begin{tabular}{ccccccccccccc}
\toprule
\multicolumn{12}{c}{\textbf{Scalars}}
\\
\midrule
\textbf{Field}&\textbf{Irrep}&\textbf{Normalization}&\textbf{Direction}&\textbf{Top}&$\bm{b\to s\ell\ell}$&$\bm{\varepsilon'/\varepsilon}$&$\bm{\delta g_Z}$&$\bm\beta$&$\bm{Q_{\sscript{W}}^{\sscript{Cs}}}$&$\bm{m_W}$&\textbf{Combined}
\\
\midrule
\multirow{2}{*}{\rule{0pt}{4ex}$\varphi\sim(\bm1,\bm2)_{\frac{1}{2}}$}
&$(\bar{\bm3}_d,\bm3_q)$
&$-|y_\varphi^d|^2/(6M_\varphi^2)$
&$\cO_{qd}^{(1)}+6\cO_{qd}^{(8)}$
&1.0 & - & 0.8 & 0.8 & - & 0.7 & 0.3 & 1.2
\\[4pt]
&$(\bar{\bm3}_q,\bm3_u)$
&$-|y_\varphi^u|^2/(6M_\varphi^2)$
&$\cO_{qu}^{(1)}+6\cO_{qu}^{(8)}$
&1.7 & 0.4 & 1.0 & 0.8 & - & 0.5 & 0.9 & 1.8
\\
\midrule
\multirow{2}{*}{\rule{0pt}{4ex}$\omega_1\sim(\bm3,\bm1)_{-\frac{1}{3}}$}
& $\bar{\bm6}_q$
&$\phantom{-}|y_{\omega_1}^{qq}|^2/(4M_{\omega_1}^2)$
&$\cO_{qq}^{(1)D}-\cO_{qq}^{(3)D}+\cO_{qq}^{(1)E}-\cO_{qq}^{(3)E}$
&1.8 & 3.6 & 0.7 & 2.9 & [1.3, 6.4] & 0.8 & 1.6 & 4.0
\\[5pt]
& $(\bar{\bm3}_d,\bar{\bm3}_u)$
&$\phantom{-}|y_{\omega_1}^{du}|^2/(3M_{\omega_1}^2)$
&$\cO_{ud}^{(1)}-3\cO_{ud}^{(8)}$
&1.1 & - & 0.8 & 0.9 & - & 0.9 & 0.4 & 1.5
\\
\midrule
$\omega_2\sim(\bm3,\bm1)_{\frac{2}{3}}$
&$\bm3_d$
&$\phantom{-}|y_{\omega_2}|^2/M_{\omega_2}^2$
&$\cO_{dd}^D-\cO_{dd}^E$
&0.4 & - & - & 0.4 & - & - & - & 0.5
\\
\multirow{1}{*}{\rule{0pt}{0ex}$\omega_4\sim(\bm3,\bm1)_{-\frac{4}{3}}$} 
&$\bm3_u$
&$\phantom{-}|y_{\omega_4}^{uu}|^2/M_{\omega_4}^2$
&$\cO_{uu}^D-\cO_{uu}^E$ 
&1.8 & - & 1.3 & 1.1 & - & 1.7 & 0.3 & 1.9
\\[3pt]
\midrule
\multirow{1}{*}{\rule{0pt}{0ex}$\zeta\sim(\bm3,\bm3)_{-\frac{1}{3}}$}
&$\bm3_q$
&$\phantom{-}|y_\zeta^{qq}|^2/(2M_\zeta^2)$
&$3\cO_{qq}^{(1)D}+\cO_{qq}^{(3)D}-3\cO_{qq}^{(1)E}-\cO_{qq}^{(3)E}$
&3.1 & 2.5 & 0.8 & 1.2 & 4.1 & 2.0 & 0.5 & 3.7
\\[3pt]
\midrule
\multirow{2}{*}{\rule{0pt}{4ex}
$\Omega_1\sim(\bm6,\bm1)_{\frac{1}{3}}$}
&$(\bm3_u,\bm3_d)$
&$\phantom{-}|y^{ud}_{\Omega_1}|^2/(6M_{\Omega_1}^2)$
&$2\cO^{(1)}_{ud}+3\cO^{(8)}_{ud}$
&1.0 & - & 0.5 & 0.8 & - & 0.9 & 0.3 & 1.4
\\[7pt]
&$\bar{\bm3}_q$
&$\phantom{-}|y_{\Omega_1}^{qq}|^2/(4M_{\Omega_1}^2)$
&$\cO_{qq}^{(1)D}-\cO_{qq}^{(3)D}-\cO_{qq}^{(1)E}+\cO_{qq}^{(3)E}$ 
&2.1 & 2.5 & 0.9 & 2.4 & [1.7, 8.3] & 1.1 & 0.6 & 2.6
\\
\midrule
$\Omega_2\sim(\bm6,\bm1)_{-\frac{2}{3}}$
&$\bm6_d$
&$\phantom{-}|y_{\Omega_2}|^2/(4M_{\Omega_2}^2)$
&$\cO_{dd}^D+\cO_{dd}^E$
&0.2 & - & - & 0.3 & - & - & - & 0.3
\\[6pt]
$\Omega_4\sim(\bm6,\bm1)_{\frac{4}{3}}$
&$\bm6_u$
&$\phantom{-}|y_{\Omega_4}|^2/(4M_{\Omega_4}^2)$
&$\cO_{uu}^D+\cO_{uu}^E$
&1.3 & 0.3 & 1.0 & 0.8 & - & 1.1 & 1.7 & 2.1
\\[6pt]
$\Upsilon\sim(\bm6,\bm3)_{\frac{1}{3}}$
&$\bm6_q$
&$\phantom{-}|y_\Upsilon|^2/(8M_\Upsilon^2)$
&$3\cO_{qq}^{(1)D}+\cO_{qq}^{(3)D}+3\cO_{qq}^{(1)E}+\cO_{qq}^{(3)E}$
&1.7 & 3.0 & 0.7 & 2.8 & 2.7 & 1.3 & 2.2 & 4.8
\\
\midrule
\multirow{2}{*}{\rule{0pt}{4ex}    
$\Phi\sim(\bm8,\bm2)_{\frac{1}{2}}$}
&$(\bar{\bm3}_q,\bm3_u)$
&$-|y^{qu}_\Phi|^2/(18M_\Phi^2)$
&$4\cO_{qu}^{(1)}-3\cO_{qu}^{(8)}$
&1.2 & 0.2 & 0.1 & 0.9 & - & 0.5 & 1.0 & 1.5
\\[4pt]
&$(\bar{\bm3}_d,\bm3_q)$
&$-|y^{dq}_\Phi|^2/(18M_\Phi^2)$
&$4\cO_{qd}^{(1)}-3\cO_{qd}^{(8)}$ 
&0.8 & - & 0.1 & 0.8 & - & 0.7 & 0.3 & 1.2
\\
\bottomrule
\end{tabular}
}
\caption{\textbf{Scalar mediators}. In the first column, we collect the labels for each mediator along with their irreps under the SM gauge group. In the second column, we list the flavor irreps, while in the third and fourth columns, we provide the normalization and the linear combination of the generated operators from the $U(3)^5$-symmetric basis (see Table~\ref{tab:MFV_basis}). In the remaining columns we collect the $95\%$ CL constraints discussed in Section~\ref{sec:EFT-limits}.}
\label{table:EFTresults_scalars}
\end{table*}

The natural next step in a systematic bottom-up approach is to construct UV completions of the $U(3)^5$-symmetric dimension-6 basis. An exhaustive leading-order classification yields a finite set of possibilities within the scope of perturbative short-distance NP. The UV/IR dictionary of Ref.~\cite{deBlas:2017xtg} is a collection of all possible SM gauge representations of new scalar, fermion, and vector fields that match onto dimensions-6 operators in the SMEFT at the tree level. Expanding on this, Ref.~\cite{Greljo:2023adz} further imposed $U(3)^5$ flavor symmetry to the UV Lagrangian, categorizing new mediator fields into irreducible representations (irreps) and defining the flavor coupling tensors. The exact symmetry limit is highly predictable --- each nontrivial flavor multiplet leads to mass degenerate states, which once integrated out, match to a single Hermitian operator in the $U(3)^5$-symmetric basis with a well-defined sign for the obtained WC.\footnote{Even for trivial flavor irreps, a notable simplification occurs, with only a few instances involving more than one parameter, see Tables 4 and 5 in~\cite{Greljo:2023adz}.} Each case predicts a direction in the WC parameter space, denoted as a \textit{leading direction}. A general tree-level matching result is a linear combination of these directions.

Needless to say, a finite number of scenarios featuring a single mediator dominance is of particular phenomenological importance. To this purpose, Ref.~\cite{Greljo:2023adz} conducted a thorough tree-level analysis for each leading direction, reporting a compendium of bounds based on the available data. This study extends the previous analysis by incorporating the RG effects. Upon a detailed case-by-case examination, we find that a substantial number of scenarios have RG-induced bounds competitive with tree-level bounds. These cases, along with their tree-level matching formulas, are presented in Table~\ref{table:EFTresults_scalars} for scalars, Table~\ref{table:EFTresults_vectors} for vectors, and Table~\ref{table:EFTresults_fermions} for fermions.

These tables explain our initial choice of operators in Table~\ref{tab:MFV_basis}. The tree-level bounds on the UV mediators which match onto the four-quark operators (scalars and vectors) are from the top quark production~\cite{Ethier:2021bye}, while those generating two-quark-two-$\phi$ operators (fermions) were constrained from a combined low-energy fit of (semi)leptonic operators~\cite{Falkowski:2017pss}. In Section~\ref{sec:EFT-limits}, we compare the tree-level with the RG-improved constraints for all mediators where the latter are numerically important.\footnote{Unlike tree-level bounds, RG-induced bounds depend on the UV renormalization scale \(\mu_i\), although this dependence is only logarithmic. In contrast, their dependence on the mass \(M_X\) and coupling \(y_X\) is quadratic, scaling as \(\propto y_X^2/M^2_X\).} In Section~\ref{sec:direct-searches}, we finally compare these indirect constraints with direct searches for two benchmark cases, revealing an interesting interplay.

\subsection{Improved EFT limits}
\label{sec:EFT-limits}

\setlength{\tabcolsep}{0.4em} 
{\renewcommand{\arraystretch}{1.4}
\begin{table*}
\centering
\scalebox{0.81}{
\begin{tabular}{cccccccccccc}
\toprule
\multicolumn{12}{c}{\textbf{Vectors}}
\\
\midrule
\textbf{Field}&\textbf{Irrep}&\textbf{Normalization}&\textbf{Direction}&\textbf{Top}&$\bm{b\to s\ell\ell}$&$\bm{\varepsilon'/\varepsilon}$&$\bm{\delta g_Z}$&$\bm\beta$&$\bm{Q_{\sscript{W}}^{\sscript{Cs}}}$&$\bm{m_W}$&\textbf{Combined}
\\
\midrule
\multirow{3}{*}{$\cB\sim(\bm1,\bm1)_0$}
&$\bm8_q$
&$-(g_\cB^q)^2/(12M_\cB^2)$
&$3\cO_{qq}^{(1)E}-\cO_{qq}^{(1)D}$
& 1.0 & 1.5 & 0.5 & 0.8 & [0.5, 2.7] & 0.4 & 0.5 & [1.4, 6.3]
\\
&$\bm8_u$
&$-(g_\cB^u)^2/(12M_\cB^2)$
&$3\cO_{uu}^E-\cO_{uu}^D$
&0.8 & 0.3 & 0.7 & 0.5 & - & 0.3 & 0.5 & 0.8
\\
&$\bm8_d$
&$-(g_\cB^d)^2/(12M_\cB^2)$
&$3\cO_{dd}^{E}-\cO_{dd}^D$
&0.2 & - & - & - & - & - & - & 0.2
\\
\midrule
$\cB_1\sim(\bm1,\bm1)_1$
&$(\bar{\bm3}_d,\bm3_u)$
&$-|g_{\cB_1}^{du}|^2/(3M_{\cB_1}^2)$
&$\cO_{ud}^{(1)}+6\cO_{ud}^{(8)}$
&1.3 & - & 1.2 & [0.3, 3.0] & - & 0.5 & - & 1.4
\\
\midrule
\multirow{1}{*}{$\cW\sim(\bm1,\bm3)_0$}
&$\bm8_q$
&$-(g_{\cW}^q)^2/(48M_{\cW}^2)$
&$3\cO_{qq}^{(3)E}-\cO_{qq}^{(3)D}$
& 0.7 & 1.6 & 0.4 & 0.5 & 1.2 & 0.5 & 0.2 & 1.9
\\
\midrule
\multirow{1}{*}{$\cQ_1\sim(\bm3,\bm2)_{\frac{1}{6}}$}
& $(\bar{\bm3}_d,\bar{\bm3}_q)$
&$2|g_{\cQ_1}^{dq}|^2/(3M_{\cQ_1}^2)$
&$\cO_{qd}^{(1)}-3\cO_{qd}^{(8)}$
& 1.3 & - & 0.9 & 0.6 & - & 0.8 & 0.2 & 1.3
\\[2pt]
\multirow{1}{*}{$ \cQ_5\sim(\bm3,\bm2)_{-\frac{5}{6}}$}
& $(\bar{\bm3}_u,\bar{\bm3}_q)$ 
&$2|g_{\cQ_5}^{uq}|^2/(3M_{\cQ_5}^2)$
&$\cO_{qu}^{(1)}-3\cO_{qu}^{(8)}$
& 2.2 & 0.3 & 1.2 & 1.0 & 0.2 & 1.2 & 0.9 & 2.1
\\[2pt]
\midrule
$\cY_1\sim(\bar{\bm6},\bm2)_{\frac{1}{6}}$ 
& $(\bar{\bm3}_d,\bar{\bm3}_q)$
&$\lzu g_{\cY_1} \dzu^2/{(3M_{\cY_1}^2)}$ 
&$2\cO_{qd}^{(1)}+3\cO_{qd}^{(8)}$ 
& 1.4 & 0.1 & 1.1 & 0.7 & - & 0.8 & 0.2 & 1.4
\\[2pt]
$\cY_5\sim(\bar{\bm6},\bm2)_{-\frac{5}{6}}$ 
& $  (\bar{\bm3}_u,\bar{\bm3}_q)$ 
&$\lzu g_{\cY_5} \dzu^2/(3M_{\cY_5}^2)$ 
&$2\cO_{qu}^{(1)}+3\cO_{qu}^{(8)}$
& 2.2 & 1.0 & 0.9 & 1.0 & 0.2 & 1.3 & 0.9 & 2.1
\\[1.5pt]
\midrule
\multirow{3}{*}{$\cG\sim(\bm8,\bm1)_0$}
& $\bm8_q$
&$-(g_\cG^q)^2/(144M_\cG^2)$ 
&$11\cO_{qq}^{(1)D}-9\cO_{qq}^{(1)E}+9\cO_{qq}^{(3)D}-3\cO_{qq}^{(3)E}$ 
& 0.9 & 0.6 & - & 0.7 & [0.7, 3.5] & 0.7 & 0.2 & 0.9
\\
& $\bm8_u$ 
&$\phantom{-}(g_\cG^u)^2/(36M_\cG^2)$
&$3\cO_{uu}^E-5\cO_{uu}^D$
& 0.7 & 0.2 & 0.4 & 0.7 & - & 0.4 & 0.3 & 0.8
\\
& $\bm8_d$ 
&$\phantom{-}(g_\cG^d)^2/(36M_\cG^2)$ 
&$3\cO_{dd}^E-5\cO_{dd}^D$
& - & - & - & - & - & - & - & [0.0, 0.4]
\\
\midrule
$\cG_1\sim(\bm8,\bm1)_1$
& $(\bar{\bm3}_d,\bm3_u)$
&$\phantom{-}\lzu g_{\cG_1} \dzu^2/(9M_{\cG_1}^2)$
&$-4\cO_{ud}^{(1)}+3\cO_{ud}^{(8)}$
& 1.1 & - & 0.2 & [0.4, 7.6] & - & 0.7 & 0.2 & 1.0
\\
\midrule
\multirow{2}{*}{$\cH\sim(\bm8,\bm3)_0$}
& $\bm8_q$ 
&$-(g_\cH)^2/(576 M_\cH^2)$
&$27\cO_{qq}^{(1)D}-9\cO_{qq}^{(1)E}-7\cO_{qq}^{(3)D}-3\cO_{qq}^{(3)E}$
& 0.4 & 0.7 & 0.2 & 0.4 & 0.8 & 0.6 & 0.2 & 0.9
\\
&$\bm1$
&$(g_\cH)^2/(96M_\cH^2)$
&$2\cO_{qq}^{(3)D}+3\cO_{qq}^{(3)E}-9\cO_{qq}^{(1)E}$
& 0.5 & 1.0 & 0.3 & 0.6 & 0.5 & - & 0.4 & 1.0
\\
\bottomrule
\end{tabular}
}
\caption{\textbf{Vector mediators}. For the description see the caption of Table~\ref{table:EFTresults_scalars}.}
\label{table:EFTresults_vectors}
\end{table*}

\setlength{\tabcolsep}{0.4em} 
{\renewcommand{\arraystretch}{1.3}
\begin{table*}
\centering
\scalebox{0.82}{
\begin{tabular}{cccccccccccc}
\toprule
\multicolumn{12}{c}{\textbf{Fermions}}
\\
\midrule
\textbf{Field}&\textbf{Irrep}&\textbf{Normalization}&\textbf{Direction}&\textbf{Top}&$\bm{b\to s\ell\ell}$&$\bm{\varepsilon'/\varepsilon}$&$\bm{\delta g_Z}$&$\bm\beta$&$\bm{Q_{\sscript{W}}^{\sscript{Cs}}}$&$\bm{m_W}$&\textbf{Combined}
\\
\midrule
\multirow{1}{*}{$U\sim(\bm3,\bm1)_{\frac{2}{3}}$}
&$\bm3_q$
&$\phantom{-}\lzu\lambda_{U}\dzu^2/(4M_{U}^2)$
&$\cO_{\phi q}^{(1)}-\cO_{\phi q}^{(3)}+\lzs 2y_u^*\cO_{u\phi}+\hermc \dzs$
&- & 2.2 & 0.6 & 5.2 & [3.1, 15.5] & 3.1 & 1.6 & 4.3
\\
\midrule
$D\sim(\bm3,\bm1)_{-\frac{1}{3}}$
&$\bm3_q$
&$-\lzu\lambda_{D}\dzu^2/(4M_{D}^2)$
&$\cO_{\phi q}^{(1)} + \cO_{\phi q}^{(3)}-\lzs 2y_d^*\cO_{d\phi}+\hermc \dzs$
&0.2 & 2.0 & 0.6 & 7.5 & [3.1, 15.5] & 2.1 & 4.0 & 6.3 
\\
\midrule
\multirow{2}{*}{$Q_1\sim(\bm3,\bm2)_{\frac{1}{6}}$}
&$\bm3_u$
&$-|\lambda^u_{Q_1}|^2/(2M_{Q_1}^2)$
&$\cO_{\phi u}-\lzs y_u^*\cO_{u\phi}+\hermc \dzs$
&- & 1.0 & 0.5 & 3.2 & 0.1 & 2.0 & 2.3 & 2.9
\\
&$\bm3_d$
&$\phantom{-}|\lambda^d_{Q_1}|^2/(2M_{Q_1}^2)$
&$\cO_{\phi d}+\lzs y_d^*\cO_{d\phi}+\hermc \dzs$
&0.2 & - & 0.5 & 3.3 & - & 3.3 & 1.2 & 4.5
\\
\midrule
\multirow{1}{*}{$Q_5\sim(\bm3,\bm2)_{-\frac{5}{6}}$}
& $\bm3_d$
&$-|\lambda_{Q_5}|^2/(2M_{Q_5}^2)$
&$\cO_{\phi d}-\lzs y_d^*\cO_{d\phi}+\hermc \dzs$
& 0.2 & - & 0.4 & 1.6 & - & 2.1 & 0.6 & [2.0, 15.4]
\\[2pt]
\multirow{1}{*}{$Q_7\sim(\bm3,\bm2)_{\frac{7}{6}}$}
& $\bm3_u$ 
&$\phantom{-}|\lambda_{Q_7}|^2/(2M_{Q_7}^2)$
&$\cO_{\phi u}+\lzs y_u^*\cO_{u\phi}+\hermc \dzs$
& - & 0.5 & 0.4 & 2.1 & - & 3.1 & 4.5 & 4.7
\\[2pt]
$T_1\sim(\bm3,\bm3)_{-\frac{1}{3}}$ 
& $\bm3_q$ 
&$\phantom{-}|\lambda_{T_1}|^2/(16M_{T_1}^2)$
&$\cO_{\phi q}^{(3)}-3\cO_{\phi q}^{(1)}+\lzs 2y_d^*\cO_{d\phi}+4y_u^*\cO_{u\phi}+\hermc \dzs$
& 0.2 & 0.6 & 0.2 & 2.0 & 3.6 & 1.8 & 3.0 & 4.2
\\[2pt]
$T_2\sim(\bm3,\bm3)_{\frac{2}{3}}$ 
& $\bm3_q$ 
&$\phantom{-}|\lambda_{T_2}|^2/(16M_{T_2}^2)$
&$\cO_{\phi q}^{(3)}+3\cO_{\phi q}^{(1)}+\lzs 4y_d^*\cO_{d\phi}+2y_u^*\cO_{u\phi}+\hermc \dzs$
& 0.1 & 0.5 & 0.2 & 2.5 & 3.6 & 2.8 & 1.6 & 3.2
\\
\bottomrule
\end{tabular}
}
\caption{\textbf{Fermion mediators}. For the description see the caption of Table~\ref{table:EFTresults_scalars}.}
\label{table:EFTresults_fermions}
\end{table*}

In this section, we derive a set of RG-improved bounds on all leading directions in the operator space of Table~\ref{tab:MFV_basis}. In Section~\ref{sec:low_energy_constraints}, the observables are calculated using the leading-log approximation to elucidate RG-induced effects. However, the results presented here rely on solving the RG equations numerically. This approach, as compared to leading-log expressions, offers several improvements. It includes effects beyond the leading log, such as  $\delta m^2_W$ from four-quark operators discussed in Section~\ref{sec:mW}, ensures resummation of large logs, and accounts for terms proportional to $g_1$ and other smaller parameters, which were omitted in the leading-log expressions but are included in numerical calculations.

We use the open-source \texttt{python} package \texttt{wilson}~\cite{Aebischer:2018bkb} to numerically solve the RG equations.\footnote{We checked that our leading-log formulae nicely agree with \texttt{wilson} where appropriate.} The UV matching scale is set at $\mu_i=3~\mathrm{TeV}$, and we run down to low energy scales, relevant for the observables discussed in Section~\ref{sec:low_energy_constraints}.\footnote{One novelty in the present analysis compared to~\cite{Greljo:2023adz} is the inclusion of the RG effects also for top quark processes, taking $\mu_f=2m_t$.} We separately construct \(\chi^2\) functions for each observable and a combined \(\chi^2\), considering both theoretical and experimental uncertainties. These functions are then used to derive 95\% CL constraints on the effective mass of each mediator, represented as a mass-to-coupling ratio. If a single number is presented, it is to be understood as a lower limit, whereas a reported interval corresponds to a $2\sigma$ preferred range.

The results for all scalar mediators are given in Table~\ref{table:EFTresults_scalars}. We observe that $b\to s\ell\ell$ plays an important role in constraining the linear combinations of $\cO_{qq}^{(1)D,E}$ and $\cO_{qq}^{(3)D,E}$ operators. This is most apparent in the cases of $\omega_1\sim\bar{\bm6}_q$ and $\Upsilon\sim\bm6_q$, where the bounds obtained using $b\to s\ell\ell$ reach $\gtrsim 3\,\mathrm{TeV}$. For the $\zeta\sim \bm3_q$ and $\Omega_1\sim\bar{\bm3}_q$, we obtain bounds which are comparable to the tree-level results, still $\gtrsim 2\,\mathrm{TeV}$. On the other hand, $\varepsilon'/\varepsilon$ bounds turn out to be less stringent, however, for $\varphi\sim(\bar{\bm3}_q,\bm3_u)$, $\omega_4\sim\bm3_u$ and $\Omega_4\sim\bm6_u$, the bounds are still around $1\,\mathrm{TeV}$. \( Z \)-pole observables impose significant bounds on the same leading directions constrained by \( b\to s\ell\ell \), extending to multi-\(\mathrm{TeV}\) ranges, and in other cases, the bounds are comparable to those derived from top data. Regarding $\beta$ decays, stringent constraints are obtained from $\cO_{qq}^{(1)D,E}$ and $\cO_{qq}^{(3)D,E}$ operators. In the case of $\zeta\sim \bm3_q$ and $\Upsilon\sim\bm6_q$, $\beta$ decays provide a highly competitive lower bound on the effective mass, while in the remaining two cases ($\omega_1\sim\bar{\bm6}_q$ and $\Omega_1\sim\bar{\bm3}_q$), we obtain a preferred range for the effective mass. The APV observable $Q_W^{\sscript{Cs}}$ yields in many cases constraints exceeding $1\,\mathrm{TeV}$, for example $\omega_4\sim\bm3_u$, $\zeta\sim \bm3_q$, and $\Upsilon\sim\bm6_q$. The modification of $m_W$, an effect beyond the leading log, is highly relevant for the $\omega_1\sim\bar{\bm6}_q$, $\Omega_4\sim\bm6_u$, $\Upsilon\sim\bm6_q$ and $\Phi\sim(\bar{\bm3}_q,\bm3_u)$ irreps, where the obtained bounds are $\gtrsim 1~\mathrm{TeV}$. In summary, the combined fit notably improves the bound for $\omega_1\sim\bar{\bm6}_q$ and $\Upsilon\sim\bm6_q$ compared to using only top data, while in other cases, it yields a modest enhancement of the bounds, although the RG-induced observables are competitive.

Numerical results for the vector mediators are presented in Table~\ref{table:EFTresults_vectors}. Highly important bounds due to $b\to s \ell\ell$ are set for $\cB\sim\bm8_q$, $\cW\sim \bm8_q$, and $\cH\sim \bm1$. $\varepsilon'/\varepsilon$ gives a bound above $1~\mathrm{TeV}$ for certain irreps, such as $\cB_1\sim(\bar{\bm3}_d,\bm3_u)$ and $\cY_1\sim(\bar{\bm3}_d,\bar{\bm3}_q)$, which are comparable to the bounds from top data. The $Z$-pole observables, $\beta$ decays, $Q_W^{\sscript{Cs}}$ and $m_W$ seldom reach the $1\, \mathrm{TeV}$ level. However, they are often still comparable to the constraints from top data alone. In some instances, $Z$-pole observables and $\beta$ decays provide a preferred range rather than a lower limit for the effective mass.

Lastly, for completeness, in Table~\ref{table:EFTresults_fermions}, we collect the constraints on fermion mediators. Tree-level bounds in Ref.~\cite{Greljo:2023adz} were derived using a combined fit of low-energy observables from Refs.~\cite{Falkowski:2017pss,Breso-Pla:2023tnz}, exceeding $3~\mathrm{TeV}$ in most cases as shown in Table 9 of Ref.~\cite{Greljo:2023adz}. Our analysis focuses on the most relevant observables from these likelihoods, presented individually. While tree-level effects primarily dominate, RG-induced constraints remain significant. As detailed in Section~\ref{sec:mW}, for UV fermion mediators, the modification of $m_W$ at leading-log order proves to be a stringent constraint in most cases, with bounds exceeding $3\,\mathrm{TeV}$ for $D\sim\bm3_q$, $Q_7\sim\bm3_u$, and $T_1\sim\bm3_q$. For $U\sim\bm3_q$ and $D\sim\bm3_q$, it is notable that $b\to s\ell\ell$, an RG-induced constraint, sets a lower limit around $2~\mathrm{TeV}$. Other constraints, largely at the tree level, are similar to those reported in Ref.~\cite{Greljo:2023adz} with minor differences.

\subsection{Direct searches}
\label{sec:direct-searches}

This subsection examines the direct search sensitivity for two heavy mediators ($\omega_1\sim \bar{\bm6}_q$ and $Q_7\sim \bm3_u$) primarily constrained at the EFT level by the RG effects discussed in this paper. We review the relevant LHC collider constraints from single and pair production for both cases.

Consider the scalar diquark $\omega_1\sim \bar{\bm6}_q$. Table~\ref{table:EFTresults_scalars} shows that for this mediator, the primary EFT constraint arises from RG mixing of $\mathcal{O}_{qq}$ into $\mathcal{O}_{\phi q}$, affecting quark FV $Z$-boson couplings and impacting $b\to s \ell \ell$ processes. The obtained constraint in the coupling versus mass plane ($\mu_i = M_{\omega_1}$) is depicted by the green shaded region in Figure~\ref{fig:omega_1_plot}, demonstrating the breaking of the pure power-law dependence on $M/y$ from the previous subsection where $\mu_i = 3\,$TeV (dashed green line). For $M_{\omega_1} = 3$\,TeV, the two approaches align, but for significantly lower or higher mass values, the discrepancies increase as expected.

The considered diquark couples to gluons and can be pair-produced in proton collisions, with a cross section set by its gauge representation and mass. Considering the ATLAS~\cite{ATLAS:2017jnp} and CMS~\cite{CMS:2018mts} searches for pair-produced colored resonances decaying to jets, we obtain a lower limit on the diquark mass of $M_{\omega_1}>700~\mathrm{GeV}$ following~\cite{Bordone:2021cca}.\footnote{Note that our diquark is a flavor multiplet with $6$ mass-degenerate states. We account for an increase of its production cross section by a factor of $6$ using the results from Ref.~\cite{Dorsner:2018ynv}.} This constraint is represented in the magenta color on Figure~\ref{fig:omega_1_plot}.
Moreover, the diquark couples to quark pairs, most notably there is a component of the flavor multiplet that couples to valence quarks.\footnote{States coupling to sea and top quarks face suppressed single production due to parton densities. QCD pair production leading to top final states~\cite{CMS:2021knz} yields bounds comparable to previously discussed jets.} This leads to important constraints from direct searches of dijet resonances from ATLAS~\cite{ATLAS:2018qto, ATLAS:2019fgd} and CMS~\cite{CMS:2018mgb, CMS:2019gwf}, which have been recast for a generic mediator in Ref.~\cite{Bordone:2021cca}. We use the results from the latter reference to obtain the constraints on Figure~\ref{fig:omega_1_plot} presented in orange. There are two notable abrupt cuts in the shown contour: firstly, the mass interval of the searches stops at $5~\mathrm{TeV}$ at most, explaining the vertical cut in the contour, and secondly, we horizontally cut the contour at the point where the partial decay width of the resonance $\Gamma_{\omega_1} = M_{\omega_1} |y_{\omega_1}^{qq}|^2/(2\pi)$ is equal to $10\%$ of its mass. Above this line, we deem the resonance to be too broad to respect the narrow-width approximation as assumed in obtaining the constraint, and only the EFT constraint applies. Ultimately, we do not consider the parameter space in which the ratio $\Gamma_{\omega_1}/M_{\omega_1}>50\%$. 

\begin{figure}[t]
     \centering
         \includegraphics[width=0.5\textwidth]{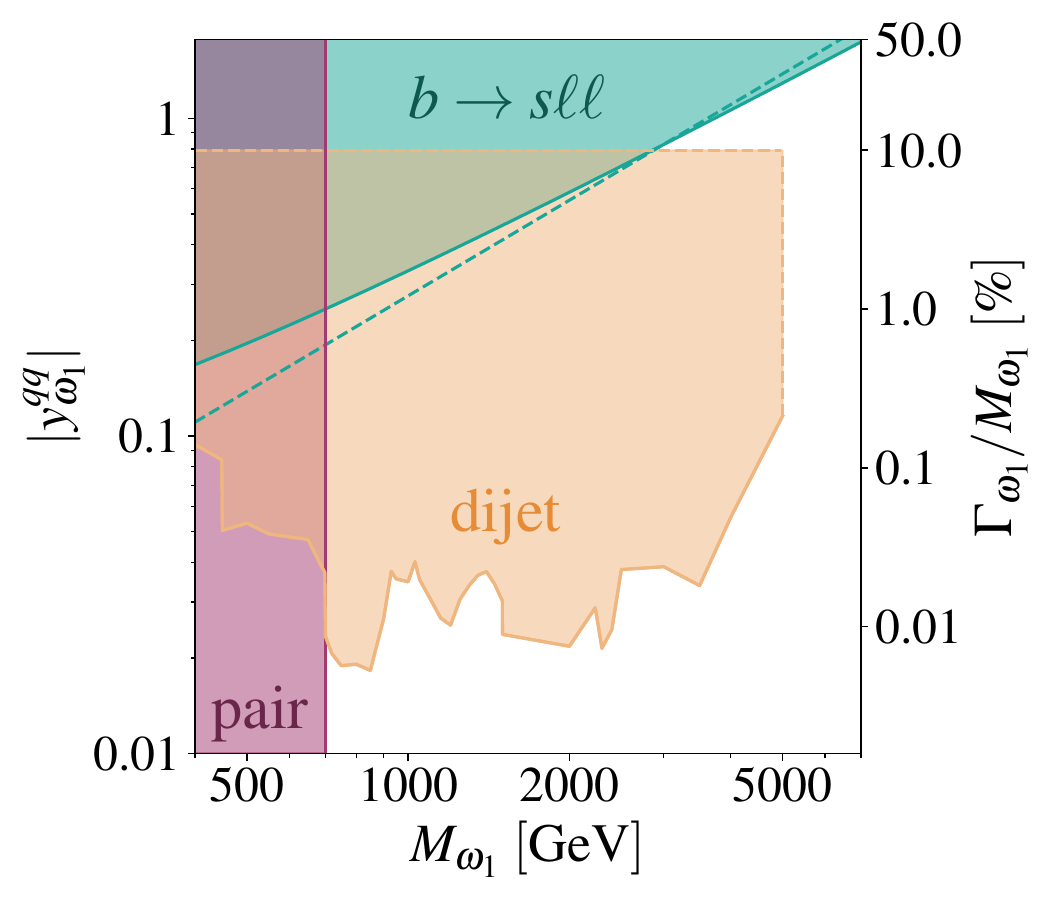}
        \caption{The leading EFT and direct searches constraints at the $95\%$ CL in the mass-coupling plane of the scalar diquark $\omega_1\sim \bar{\bm6}_q$. See Section~\ref{sec:direct-searches} for details.}
        \label{fig:omega_1_plot}
\end{figure}

The limited parameter space permitted by $b \to s \ell \ell$ and not excluded by direct searches, visible in the upper part of Figure~\ref{fig:omega_1_plot}, could be explored through broad (and heavy) resonance searches. This coincides with the region of interest for contact interaction searches in dijet production, which utilize angular distributions~\cite{ATLAS:2017eqx, CMS:2018ucw}. Regrettably, the constraints on the WCs obtained in these studies are not directly transferable to our work because the dijet invariant mass underlying these bounds lies in the multi-TeV range, where a mediator-based description is more suitable. A further challenge is that ATLAS and CMS analyze a narrow range of operators, unsuitable even within the restrictive $U(3)^3$ flavor structure. We recommend that experimental collaborations adjust future dijet data interpretations accordingly.

We analyze the vector-like quark $Q_7\sim \bm3_u$ for our second example. As Table~\ref{table:EFTresults_fermions} indicates, the most significant EFT constraint is the RG-induced shift in the $W$-boson mass, imposing a lower limit of over $ 4.5~\mathrm{TeV} $ on the mass-to-coupling ratio. Concurrently, LHC searches for pair-produced vector-like quarks, particularly ATLAS, have established a lower bound of  $1.3~\mathrm{TeV} $ for top quark partners~\cite{ATLAS:2018ziw}. This limit applies to the mass of $Q_7$ since part of its multiplet interacts with the top quark. It is worth noting that single production searches for third-generation vector-like quarks~\cite{CMS:2022yxp} offer similar, though slightly less stringent probe, due to the parton density suppression. A dedicated search for first-generation vector-like quarks would be beneficial.

\section{Conclusion}
\label{sec:conclusions}

This paper explores the intricate phenomenology emerging from renormalization group equations in the SMEFT. We rigorously assess the most significant RG mixing patterns by focusing on microscopic theories whose dominant effects are captured with a $U(3)^5$-symmetric dimension-6 operator basis at the UV scale. A key finding is that RG-induced effects on low-energy precision observables often lead to constraints on four-quark (and two-quark) operators that rival or surpass those derived from tree-level processes, notably in top-quark physics. We thoroughly investigate a wide array of single mediator models that match onto the $U(3)^5$-symmetric basis at the tree-level, identified as \textit{leading directions}. We provide a comprehensive RG analysis of these directions, extending beyond the earlier tree-level study~\cite{Greljo:2023adz}. The key outcomes of this analysis are detailed in three tables: Table~\ref{table:EFTresults_scalars} for scalar mediators, Table~\ref{table:EFTresults_vectors} for vectors, and Table~\ref{table:EFTresults_fermions} for fermions.

Moving forward, our numerical analysis indicates that the observables we have examined are probing tree-level physics at the TeV scale, characterized by order one couplings, where direct searches can offer complementary probes, as depicted in Figure~\ref{fig:omega_1_plot}. A pivotal aspect of future work will be the thorough examination of dijet data within the SMEFT framework and across explicit mediator models, improving the direct search strategies for leading directions, as well as quantifying the validity of the EFT interpretation. Another promising direction is the exploration of $U(2)^5$ flavor symmetry in a similar context. This lower symmetry increases the set of tree-level mediators, adding more complexity and perhaps offering a better benchmark for the physics lying beyond the SM.

\begin{acknowledgments}

This work received funding from the Swiss National Science Foundation (SNF) through the Eccellenza Professorial Fellowship ``Flavor Physics at the High Energy Frontier'' project number 186866.

\end{acknowledgments}

\bibliography{letter}

\end{document}